\documentclass[12pt]{iopart}
\usepackage{graphicx}
\usepackage{epsfig}
\begin{document}


\title{From a 1D completed scattering and double slit diffraction to the
quantum-classical problem: A new approach}

\author{N L Chuprikov}

\address{Tomsk State Pedagogical University, 634041, Tomsk, Russia}
\begin{abstract}

We present a new approach to the quantum-classical problem, which treats it as the
problem of modelling the quantum phenomenon described by a coherent superposition of
microscopically distinct substates (CSMDS) as a compound one consisting of
alternative subprocesses creating unremovable contexts for each other, or as that of
reducing a non-Kolmogorovian quantum probability space to underlie a CSMDS to the
sum of Kolmogorovian ones. We develop such models for a 1D completed scattering and
double slit diffraction. The quantum-classical problem disappears when, in quantum
theory with its integral superposition principle, CSMDSs obey the "either-or" rule
to guide alternative random events. There is no observable which could be associated
with the whole ensemble of statistical data described by a CSMDS, because such data
are incompatible -- in the case of a CSMDS, any observable splits into noncommuting
observables associated with the substates. To calculate the average value of any
observable as well as to introduce characteristic times is meaningful only for the
substates of a CSMDS. Ignoring this feature in the conventional description of
CSMDSs just leads to paradoxical results (e.g., to the Hartman effect and passing a
particle through two slits in the screen simultaneously).

\end{abstract}


\maketitle

\newcommand{\Api}{A^{in}}
\newcommand{\Ami}{B^{in}}
\newcommand{\Apo}{A^{out}}
\newcommand{\Amo}{B^{out}}
\newcommand{\bpi}{a^{in}}
\newcommand{\bmi}{b^{in}}
\newcommand{\bpo}{a^{out}}
\newcommand{\bmo}{b^{out}}
\newcommand{\api}{a^{in}}
\newcommand{\ami}{b^{in}}
\newcommand{\apo}{a^{out}}
\newcommand{\amo}{b^{out}}
\newcommand {\uta} {\tau_{tr}}
\newcommand {\utb} {\tau_{ref}}

\section{Introduction} \label{intr}

A 1D completed scattering and double slit diffraction represent the most simple
one-particle scattering problems to arise in quantum mechanics, and their
conventional quantum-mechanical models are commonly accepted to give an exhaustive,
maximally possible in quantum mechanics, description. The only reason to compel
physicists to address these phenomena again and again is an irresistible urge to
explain a counterintuitive character of particle's properties to follow from these
models: by them a particle may tunnel with a superluminal velocity through an opaque
potential barrier and pass simultaneously through two slits in the screen.

However, despite these paradoxical properties, in the literature nobody calls in
question the models themselves. And an important role in this case is played by the
fact that they are consistent with Bell's analysis of the thought
Einstein-Podolsky-Rosen-Bohm (EPR-Bohm) experiment, from which it follows the
nonexistence of local hidden (objective, predetermined) variables and, on the
contrary, the existence of nonzero quantum correlations between events separated by
a spatial-like interval. Against this background the above properties of a
(micro)particle look as its inherent ones.

So far the EPR-Bohm experiment and these two one-particle phenomena are at the heart
of the long-standing debates on the problem of the quantum-to-classical transition
(quantum-classical problem) and foundations of quantum mechanics (see. e.g.,
\cite{Leg}). Their common feature is that in all the cases we meet pure quantum
states to represent coherent superpositions of microscopically distinct substates
(CSMDSs) whose conventional description prevents the quantum-to-classical transition
when one tries to keep the quantum-mechanical superposition principle as an
universal law, both at the micro- and macro-levels (the Schr\"odinger's cat
paradox).

In fact the conventional treatment of CSMDSs and observed violation of Bell's
inequalities compel investigators to search for solving the quantum-classical
problem beyond the idealization of isolated system, by assuming that there is some
unremovable, external for the studied system factor to suppress the action of the
superposition principle at the macro-level and, thereby, to transform its
time-dependent CSMDS into some substate of the CSMDS.

However, by the following reasons this programm raises objections. Indeed, this
programm dooms quantum theory to be unable to {\it explain} such phenomena -- within
it quantum phenomena described by CSMDSs remain "unspeakable" (in Bell's terminology
\cite{Bell}). Besides, within it quantum mechanics, with its superposition
principle, looses its status of a universal theory valid both on the micro- and
macro-scales. At the same time Bell himself considered that "The quantum phenomena
do not exclude a uniform description of micro and macro worlds\dots"\ \cite{Bell}.
In answering the question "wave {\it or} particle?", he said, together with de
Broglie, "wave {\it and} particle" [ibid].

Our aim in this paper is to present an alternative programm of solving the
quantum-classical problem, which renders "speakable"\ the micro-world, with keeping
the idealization of isolated systems and initial status of quantum mechanics as an
universal theory. This programm is based on the studies of Bell's inequalities and
CSMDSs, from the viewpoint of classical probability theory (see, respectively,
review \cite{Hre} and paper \cite{Ac1}), as well as on the developed in
\cite{Ch26,Ch27,Ch3,Ch29} novel quantum-mechanical approach to a 1D completed
scattering and a double slit diffraction, which treats both these one-particle
processes as compound ones, i.e., it treats them in the spirit of classical physics
where a particle can be {\it either} transmitted {\it or} reflected by the potential
barrier and cannot pass through two slits simultaneously.

These approaches show that the conventional treatment of CSMDSs and experimental
violation of Bell's inequalities cannot be considered as a finally established fact.
As is shown in \cite{Ch26,Ch27,Ch3,Ch29}, the "either-or"\ rule to guide mutually
exclusive random events, {\it must} and {\it can} be extended onto the micro-level,
i.e., onto CSMDSs. This step allows one both to overcome the interpretational
problems surrounding CSMDSs and consistently define characteristic times for the
particle's dynamics described by such states. The latter important, because the
conventional description of CSMDSs leads to anomalously short or even negative
values of the tunnelling time.

Note that within the conventional description of CSMDSs, where their properties are
"unspeakable", all interpretational problems are reduced to the question of the
unambiguous interpretation of the corresponding experimental data. In this
connection, it is relevant to quote from Bohr \cite{Bohr} who wrote, regarding the
foundations of quantum mechanics, that "the unambiguous interpretation of any
measurement must be essentially framed in terms of the classical physical theories"
or, else, "it is decisive to recognize that, however far the phenomena transcend the
scope of classical physical explanation, the account of all evidence must be
expressed in classical terms".

As is known, this requirement have led Bohr to the complementarity principle by
which, in particular, the wave and corpuscular properties of a particle cannot be
observed simultaneously. However, this requirement is only a part of truth, because
a consistent interpretation of any experiment must be also based on that theory to
describe the phenomenon under study (and the interpretation of the experimental
violation of Bell's inequalities is a representative example). This means that the
unambiguous interpretation of statistical experimental data obtained in studying
quantum phenomena can be reached if only quantum theory respects classical
probability theory.

In particular, the {\it unambiguous} interpretation of experiments associated with
CSMDSs is realizable if only quantum mechanics (with its integral superposition
principle) respects the "either-or"\ rule to guide incompatible events. Since the
contemporary quantum theory of CSMDSs does not obey this requirement, the
complementarity principle, based essentially on this theory, does not reflect the
inherent nature of the wave-particle duality. The models \cite{Ch26,Ch27,Ch3,Ch29}
show that in the processes under study a particle exhibits simultaneously both
corpuscular and wave properties, which respect each other. (We have to stress that
these models call in question only the complementarity of the wave and corpuscular
properties of a particle; they do not touch the validity of other aspects of the
complementarity principle.)

The plan of the paper is as follows. In Sections \ref{a0} we show that the
interpretational problems to appear at present in studying the temporal aspects of
tunnelling result from the fact that the conventional model of a 1D completed
scattering does not allow in principle a consistent resolution of the tunnelling
time problem (TTP). A new model of this process and solving the TTP on its basis are
presented in Section \ref{a1}. In Section \ref{slit} we revise the model of a double
slit diffraction. In Section \ref{a4} we discuss solving the quantum-classical
problem from the viewpoint of these two models.

\section{On the impossibility of a consistent solving of the TTP within the conventional
model of a 1D completed scattering} \label{a0}

As is known, the quantum-mechanical model of tunnelling a particle through a
one-dimensional static potential barrier (hereinafter referred to as "conventional
model of tunnelling" (CMT)) has been included in many textbooks on quantum mechanics
as the representative of an exhaustive description of quantum phenomena. However,
studying the temporal aspects of a 1D completed scattering (see reviews
\cite{Ha2,La1,Olk,Ste,Mu0,Nu0,Ol2,Win} and references therein) showed that the
simplicity of this process is illusive.

Now it is well known that the CMT predicts the Hartman effect -- for a particle
tunnelling through a single opaque potential barrier the tunnelling time saturates
with increasing the barrier's width -- the (usual) Hartman effect. Moreover, for a
particle tunnelling through several successive barriers, the tunnelling time
saturates even with increasing the space between the barriers -- the generalized
Hartman effect. Both these properties of the tunnelling time concept introduced in
the CMT say about a superluminal effective velocity of a tunnelling particle. And
what is important is that the Hartman effect has been found both in the case of the
(nonrelativistic) Schr\"odinger equation (see, e.g.,
\cite{Har,But,Mu6,Wi1,Ol1,Ol2,So1,Rec}) and the (relativistic) Dirac equation (see,
e.g., \cite{Krek,Zhi,Lun}). It is evident that these paradoxical results need a
proper explanation.

Deep controversy raised by the Hartman effect has not yet been overcome. All
attempts to reconcile this prediction of the CMT with special relativity and, thus,
to justify the existing tunnelling time concepts as those to characterize a {\it
particle}, have not been successful. In this connection, it is worthwhile to point
here to the concluding diagnoses made by Winful and Nimtz: by Winful "\ldots the
group delay in tunnelling is not a transit time \ldots" \cite{Wi1}; by Nimtz "\ldots
tunnelling modes propagate in zero time. They arise via virtual particles"
\cite{Nim}. Summing up both these statements we conclude that so far there is no
consistent definition of the tunnelling (transit) time for real (not virtual)
particles.

In \cite{Wi1} Winful points to the possible ways to overcome the interpretational
problems associated with the Hartman effect. In particular, he says that "\dots the
duration of the tunnelling event will simply be the temporal extent of the wave
packet, assumed propagating at its initial velocity" \cite{Wi1}. However, it is
difficult to agree with this statement. Yes, this {\it ad hoc} definition of "the
duration of the tunnelling event"\ does not lead to the Hartman effect. However, it
is incorrect to equate a particle with the (spreading) wave packet to describe its
state. As well as it is inconsistent to determine the {\it tunnelling} time on the
basis of the wave packet to describe the whole ensemble of particles, including {\it
reflected} particles.

This also concerns another, referred in \cite{Wi1}, "luminal"\ characteristic time
for a tunnelling particle -- "the net-flux delay"\ -- which is introduced "by
dividing [the] dwell time by the transmission coefficient". Again, the Hartman
effect disappears in this case. However, the renormalization {\it by hand} cannot
legalize the time quantity to describe the whole ensemble of particles,
to-be-transmitted and to-be-reflected, as that to describe only {\it tunnelling}
particles.

What is the reason to prevent consistent definition of the tunnelling (or transit)
time within the CMT? The answer is that "\ldots an incoming peak or centroid does
not, in any obvious physically causative sense, turn into an outgoing peak or
centroid\ldots" (see \cite{La2} as well as \cite{La1,Wi1}). That is, the CTM does
not provide the knowledge of the time evolution of transmitted particles at all
stages of scattering, which is needed for a consistent definition of the tunnelling
(transit) time. It is evident that it can be done only within the model to treat a
1D completed scattering as a compound process consisting of two subprocesses --
transmission and reflection.

As was shown in \cite{Ch26,Ch27,Ch3,Ch29}, the standard Schr\"odinger equation
allows one to develop such a model. All characteristic times introduced in
\cite{Ch26,Ch27,Ch3,Ch29} are in a full agreement with special relativity. Let us
show this in details.

\newcommand{\ko}{\kappa_0^2}
\newcommand{\kj}{\kappa_j^2}
\newcommand{\kd}{\kappa_j d_j}
\newcommand{\kki}{\kappa_0\kappa_j}

\newcommand{\Ra}{R_{j+1}}
\newcommand{\Rb}{R_{(1,j)}}
\newcommand{\Rc}{R_{(1,j+1)}}

\newcommand{\Ta}{T_{j+1}}
\newcommand{\Tb}{T_{(1,j)}}
\newcommand{\Tc}{T_{(1,j+1)}}

\newcommand{\Wa}{w_{j+1}}
\newcommand{\Wb}{w_{(1,j)}}
\newcommand{\Wc}{w_{(1,j+1)}}

\newcommand{\UU}{u^{(+)}_{(1,j)}}
\newcommand{\VV}{u^{(-)}_{(1,j)}}

\newcommand{\ta}{t_{j+1}}
\newcommand{\tb}{t_{(1,j)}}
\newcommand{\tc}{t_{(1,j+1)}}

\newcommand{\tee}{\vartheta_{(1,j)}}

\newcommand{\tta}{\tau_{j+1}}
\newcommand{\ttb}{\tau_{(1,j)}}
\newcommand{\ttc}{\tau_{(1,j+1)}}

\newcommand{\FF}{\chi_{(1,j)}}
\newcommand {\aro}{(k)}
\newcommand {\da}{\partial}
\newcommand{\ppp}{\mbox{\hspace{5mm}}}
\newcommand{\ooo}{\mbox{\hspace{3mm}}}
\newcommand{\ooa}{\mbox{\hspace{1mm}}}
\newcommand{\ppa}{\mbox{\hspace{2cm}}}

\section{A new quantum-mechanical model of a 1D completed scattering} \label{a1}
\subsection{Backgrounds} \label{a10}

A 1D completed scattering is considered in \cite{Ch26,Ch27,Ch3} in the following
setting. A particle impinges a symmetrical potential barrier $V(x)$
($V(x-x_c)=V(x_c-x)$) confined to the finite spatial interval $[a,b]$ $(a>0)$;
$d=b-a$ is the barrier width, the point $x_c$ is the midpoint of the barrier region.
At the initial instant of time, long before the scattering event, the state of a
particle $\Psi_{full}^{(0)}(x)$ approaches the in-asymptote
\[\fl\Psi_{full}^{in}(x,t)=\frac{1}{\sqrt{2\pi}}\int_{-\infty}^{\infty}\Api(k)\exp[i(kx-
E(k)t/\hbar)]dk,\] which is supposed to be a normalized function to belong to the
set $S_{\infty}$ consisting from infinitely differentiable functions vanishing
exponentially in the limit $|x|\to \infty$; $E(k)=\hbar^2k^2/2m$. Without loss of
generality, it is also supposed that
\begin{eqnarray} \label{444}
\fl<\Psi_{full}^{(0)}|\hat{x}|\Psi_{full}^{(0)}>=0,\ooo
<\Psi_{full}^{(0)}|\hat{p}|\Psi_{full}^{(0)}> =\hbar k_0
> 0,\ooo <\Psi_{full}^{(0)}|\hat{x}^2|\Psi_{full}^{(0)}> =l_0^2,
\end{eqnarray}
where $l_0$ and $k_0$ are given parameters ($l_0<<a$); $\hat{x}$ and $\hat{p}$ are
the operators of the particle's position and momentum, respectively. For the
Gaussian wave packet $\Api(k)=(2l_0^2/\pi)^{1/4} \exp[-l_0^2(k-k_0)^2]$. For a
completed scattering the average velocity $\hbar k_0/m$ of incident particles, i.e.,
the velocity of the centroid of the incident wave packet, is supposed to be much
more than the rate of its spreading.

For any value of $t$ the wave function to describe the particle's state has the form
\begin{eqnarray} \label{11}
\fl\Psi_{full}(x,t)=\frac{1}{\sqrt{2\pi}}\int_{-\infty}^{\infty}
\Api(k)\Psi_{full}(x;k)\exp[-i E(k)t/\hbar]dk;
\end{eqnarray}
where $\Psi_{full}(x;k)$, the stationary state of a particle, can be written as
follows
\begin{eqnarray} \label{511}
\fl \Psi_{full}(x;k)=\left\{ \begin{array}{c} e^{ikx}+b_{out}(k)e^{ik(2a-x)}\ppp (x\le a);\\
a_{full}\cdot u(x-x_c;k)+b_{full}\cdot v(x-x_c;k) \ppp (a\le x\le b);\\
a_{out}(k)e^{ik(x-d)}\ppp (x>b);
\end{array}\right.
\end{eqnarray}
$u(x-x_c;k)$ and $v(x-x_c;k)$ are such real solutions to the Schr\"odinger equation
that $u(x_c-x;k)=-u(x-x_c;k)$, $v(x_c-x;k)=v(x-x_c;k)$;
$\frac{du}{dx}v-\frac{dv}{dx}u=\kappa$ is a constant;
\begin{eqnarray} \label{51300}
\fl a_{out}=\frac{1}{2}\left(\frac{Q}{Q^*}-\frac{P}{P^*}\right);\ooo
b_{out}=-\frac{1}{2}\left(\frac{Q}{Q^*}+\frac{P}{P^*}\right);
\end{eqnarray}
\begin{eqnarray*}
\fl a_{full}=\frac{1}{\kappa}\left(P+P^*b_{out}\right)e^{ika}=
-\frac{1}{\kappa}P^*a_{out}e^{ika};\ooo
b_{full}=\frac{1}{\kappa}\left(Q+Q^*b_{out}\right)e^{ika}=
\frac{1}{\kappa}Q^*a_{out}e^{ika};\nonumber
\end{eqnarray*}
\begin{eqnarray*}
\fl Q=\left(\frac{du(x-x_c)}{dx}+i k u(x-x_c)\right)\Bigg|_{x=b};\ooa
P=\left(\frac{dv(x-x_c)}{dx}+i k v(x-x_c)\right)\Bigg|_{x=b}.\nonumber
\end{eqnarray*}

Note, in the case of the rectangular barrier of height $V_0$ we have
\begin{eqnarray} \label{6}
\fl u=\sinh(\kappa x),\ooo v=\cosh(\kappa x),\ooo \kappa=\sqrt{2m(V_0-E)}/\hbar\ooo
(E<V_0); \nonumber\\
\fl u=\sin(\kappa x),\ooo v=\cos(\kappa x),\ooo\kappa=\sqrt{2m(E-V_0)}/\hbar\ooo
(E\geq V_0).
\end{eqnarray}

\subsection{Searching-for the incoming waves causally connected to
the transmitted and reflected ones} \label{a11}

Now we can proceed to the crucial step of our approach -- to finding the incoming
waves connected causally to the outgoing ones both for transmission and reflection.
For this purpose we have to formulate the first two physical requirements on the
searched-for stationary waves: (1) a causal relationship between the incoming wave
and the outgoing wave implies the continuity of the wave function to describe each
subprocess, as well as the continuity of the corresponding probability current
density; (2) the superposition of the incoming waves for transmission and reflection
must give the incoming wave to describe the whole scattering process.

In accordance with these requirements our aim now is to find two solutions to the
Schr\"odinger equation, $\Psi_{tr}(x;k)$ and $\Psi_{ref}(x;k)$, such that the
amplitudes of the incoming wave of $\Psi_{tr}(x;k)$ ($\Psi_{ref}(x;k)$) and the
transmitted (reflected) wave of $\Psi_{full}(x;k)$ are equal by modulus, and,
besides, $\Psi_{tr}(x;k)+\Psi_{ref}(x;k)=\Psi_{full}(x;k)$. Thus, for $x\leq a$
\begin{eqnarray} \label{700}
\fl\Psi_{tr}(x;k)=\Api_{tr}e^{ikx}, \ppp
\Psi_{ref}(x;k)=\Api_{ref}e^{ikx}+b_{out}e^{ik(2a-x)}
\end{eqnarray}
where $\Api_{tr}+\Api_{ref}=1$; $|\Api_{tr}|=|a_{out}|$, $|\Api_{ref}|=|b_{out}|$.

As was shown in \cite{Ch26}, there are two pair of solutions to the Schr\"odinger
equation, whose incoming waves obey these requirements. In one pair
$\Psi_{ref}(x;k)$ is an even function, relative to the midpoint $x_c$; but in
another pair, it is an odd one. As it will be seen from the following, only the last
pair of solutions is associated with the subprocesses. For this pair
\begin{eqnarray*}
\fl\Api_{ref}=b_{out}\left(b^*_{out}-a^*_{out}\right)\equiv
b_{out}^*\left(b_{out}+a_{out}\right);\ooa
\Api_{tr}=a^*_{out}\left(a_{out}+b_{out}\right)\equiv
a_{out}\left(a^*_{out}-b^*_{out}\right).
\end{eqnarray*}
We have to stress that not only $\Api_{tr}+\Api_{ref}=1$, but also
$|\Api_{tr}|^2+|\Api_{ref}|^2=1$. In terms of the (real) transmission and reflection
coefficients, $T(k)$ and $R(k)$, we have $\Api_{ref}=\sqrt{R}(\sqrt{R}\pm i\sqrt{T})
\equiv \sqrt{R}\exp(i\lambda)$, $\Api_{tr}=\sqrt{T}(\sqrt{T}\mp i\sqrt{R}) \equiv
\sqrt{T}\exp\left[i\left(\lambda+sign(\lambda)\frac{\pi}{2}\right)\right]$;
$\lambda=\pm\arctan(\sqrt{T/R})$; $T=|a_{out}|^2$, $R=|b_{out}|^2$. For these
amplitudes
\begin{eqnarray} \label{701}
\fl\Psi_{ref}(x;k)=\left\{ \begin{array}{c} \kappa^{-1}\left(PA^{in}_{ref}+
P^*b_{out}\right)e^{ika}u(x-x_c;k)\ppp (a\le x\le b);\\
-b_{out}e^{ik(x-d)}+\Api_{ref}e^{ik(2x_c-x)} \ppp (x\ge b);
\end{array}\right.\nonumber\\
\fl\Psi_{tr}(x;k)=\left\{ \begin{array}{c} \kappa^{-1}PA^{in}_{tr}e^{ika}u(x-x_c;k)+
b_{ful}v(x-x_c;k)\ppp (a\le x\le b);\\
(a_{out}+b_{out})e^{ik(x-d)}-\Api_{ref}e^{ik(2x_c-x)} \ppp (x\ge b)
\end{array}\right.
\end{eqnarray}

So, the solutions $\Psi_{tr}(x;k)$ and $\Psi_{ref}(x;k)$ aimed to describe the
subprocesses are determined by Exps. (\ref{700}) and (\ref{701}). As is seen, both
the solutions contain waves to impinge the barrier from the right and return
backward, without crossing the point $x_c$. So that, unlike $\Psi_{full}(x;k)$, each
of the solutions $\Psi_{tr}(x;k)$ and $\Psi_{ref}(x;k)$ describes the scattering
problem with two sources of particles. Thus, they themselves cannot describe the
subprocesses in the original problem where there is only one source of particles.

\subsection{Wave functions for transmission and reflection} \label{a12}

Note that the "extra" waves in the region $x>x_c$ disappear in the superposition
$\Psi_{tr}(x;k)+\Psi_{ref}(x;k).$ Moreover, the incident wave $\Api_{ref}e^{ikx}$
describes particles which do not cross the point $x_c$, as $\Psi_{ref}(x_c;k)=0$ for
all values of $k$. Therefore, in the region $x<x_c$, the only causal counterpart to
the transmitted wave is the incident wave $\Api_{tr}e^{ikx}$.

Thus, for symmetrical barriers, the midpoint of the barrier region is a particular
point for the subprocesses. It divides the $OX$-axis into two parts where they are
described by different solutions of the Schr\"odinger equation. Transmission and
reflection are described, respectively, by the functions $\psi_{tr}(x;k)$ and
$\psi_{ref}(x;k)$:
\begin{eqnarray} \label{2}
\fl\psi_{ref}(x;k)\equiv \Psi_{ref}(x;k),\ppp \psi_{tr}(x;k)\equiv
\Psi_{tr}(x;k)\ppp (x\le x_c);\nonumber\\
\fl\psi_{ref}(x;k)\equiv 0,\ppp \psi_{tr}(x;k)\equiv \Psi_{full}(x;k)\ppp (x\ge
x_c).
\end{eqnarray}
The main peculiarity of $\psi_{tr}(x;k)$ and $\psi_{ref}(x;k)$ is that each of them
contains one incoming and one outgoing wave. As is seen from (\ref{2}), despite the
fact that either is presented, in the regions $x<x_c$ and $x>x_c$, by different
solutions of the Schr\"odinger equation, these functions as well as the
corresponding probability current densities are continuous at the point $x_c$. Note
that the rejected even solution $\Psi_{ref}(x;k)$ does not lead to continuous wave
functions for subprocesses.

So, by this approach, reflected particles never cross the point $x_c$ in the course
of scattering. This result agrees with the well known fact that, for a classical
particle to impinge from the left a smooth symmetrical potential barrier, the
midpoint of the barrier region is the extreme right turning point, irrespective of
the particle's mass and the barrier's form and size. By the terminology of
\cite{Hr1} this fact means that in the regions $x<x_c$ and $x>x_c$ a particle moves
under different experimental contexts. In a sense, this explains why transmitted
particles are described by the wave function $\psi_{tr}(x;k)$ represented in these
regions by (though properly matched but) different solutions of the Schr\"odinger
equation -- different contexts imply different time evolutions of the ensemble.

For narrow in $k$-space wave packets, i.e., in the limit $l_0\to\infty$, the wave
packet $\Psi_{full}(x,t)$ (see (\ref{11})) as well as the ones $\psi_{tr}(x,t)$ and
$\psi_{ref}(x,t)$ formed respectively from $\psi_{tr}(x;k)$ and $\psi_{ref}(x;k)$
obey the following relations
\[\fl\Re\langle\psi_{tr}(x,t)|\psi_{ref}(x,t)\rangle=\int_{-\infty}^{x_c}
\Re\left[\Psi_{tr}^*(x,t)\Psi_{ref}(x,t)\right]dx=0.\] Therefore, despite the
existence of interference between $\psi_{tr}$ and $\psi_{ref}$, for any $t$ we have
\begin{eqnarray} \label{3}
\fl \langle\Psi_{full}(x,t)|\Psi_{full}(x,t)\rangle
=\textbf{T}+\textbf{R}=1;\nonumber\\
\fl \textbf{T}=\langle\psi_{tr}(x,t)|\psi_{tr}(x,t)\rangle=\int_{-\infty}^{x_c}
\left|\Psi_{tr}(x,t)\right|^2dx+ \int^{\infty}_{x_c}
\left|\Psi_{full}(x,t)\right|^2dx;\\
\fl \textbf{R}=\langle\psi_{ref}(x,t)|\psi_{ref}(x,t)\rangle=\int_{-\infty}^{x_c}
\left|\Psi_{ref}(x,t)\right|^2dx;\nonumber
\end{eqnarray}
constants $\textbf{T}$ and $\textbf{R}$ are the transmission and reflection
coefficients, respectively.

Eqs. (\ref{3}) just support the idea that a 1D completed scattering can be presented
as a compound process consisting of two alternative subprocesses -- transmission and
reflection. These subprocesses are inseparable from each other, because either
subprocess creates an unremovable context for its counterpart.

Note, for wave packets of any width, $\textbf{R}$ remains unchanged at all stages of
scattering. However $\textbf{T}$ is now constant only at the initial and final
stages, i.e., long before and long after the scattering event:
$\textbf{T}=\int_{-\infty}^\infty |A_{tr}^{in}(k)|^2T(k)dk=\int_{-\infty}^\infty
|a_{out}(k)|^2T(k)dk=1-\textbf{R}$. At the very stage of scattering,
$d\textbf{T}/dt=I_{tr}(x_c+0,t)-I_{tr}(x_c-0,t)\neq 0$; here $I_{tr}$ is a
probability current density to correspond to $\psi_{tr}(x,t)$.

Thus, in the general case, for times to correspond to the scattering event, the
quantum mechanical formalism does not allow one to entirely exclude the interference
terms from $\psi_{tr}(x,t)$, in partitioning the whole scattering process into
alternative subprocesses.  It should be stressed however that, in our numerical
calculations for wave packets whose initial width was comparable with the barrier
width, the relative deviation of $\textbf{T}$ from $1-\textbf{R}$ did not exceed
several percentages.

Of course, the alternation of the Schr\"odinger evolution of the subprocesses at the
point $x_c$ leads also to other peculiarities of the subprocesses. Let us consider
in detail the case of narrow in $k$-space wave packets when the variation of the
norm $\textbf{T}$ is negligible. For example, of interest is the fact that, for such
packets, reflected particles are affected at the point $x_c$ by an extra average
force to push particles out from the barrier region, backward into the left
out-of-barrier region -- \[\fl
\frac{d<\hat{p}>_{ref}}{dt}=\left<-\frac{dV}{dx}\right>_{ref}-
\frac{\hbar^2}{2m}\left|\frac{\partial\psi_{ref}}{\partial x}\right|^2_{x=x_c-0};\]
here angle brackets denote averaging over the corresponding ensemble of particles.
For transmitted particles, in the analogous expression
\[\fl \frac{d<\hat{p}>_{tr}}{dt}=\left<-\frac{dV}{dx}\right>_{tr}+
\frac{\hbar^2}{2m}\left(\left|\frac{\partial\psi_{tr}}{\partial
x}\right|^2_{x=x_c+0}-\left|\frac{\partial\psi_{tr}}{\partial
x}\right|^2_{x=x_c-0}\right),\] the second term equals to zero. Indeed, in the limit
$l_0\to\infty$, we have
\[\fl \left|\frac{\partial\psi_{tr}}{\partial
x}\right|^2_{x=x_c+0}-\left|\frac{\partial\psi_{tr}}{\partial
x}\right|^2_{x=x_c-0}=\kappa^2\left(|a_{full}|^2-|a_{tr}|^2\right)=0,\] because the
coefficients $a_{full}(k)$ and $a_{tr}(k)$ are equal by module (see (\ref{51300})).

What is important for introducing the group scattering times is that, in this
limiting case, the alternation of the Schr\"odinger evolution of the subprocesses at
the point $x_c$ does not lead to extra terms in the time derivatives for the
particle's position $x$: \[\fl
\frac{d<\hat{x}>_{tr}}{dt}=\frac{1}{m}\left<p\right>_{tr} \ppp
\frac{d<\hat{x}>_{ref}}{dt}=\frac{1}{m}\left<p\right>_{ref}.\]

Our next step is to present, in the limit $l_0\to\infty$, characteristic times for
the subprocesses of a 1D completed scattering.

\subsection{Characteristic times for transmission and reflection} \label{a13}
\subsubsection{Exact and asymptotic group times for transmission and
reflection} \label{a21}

A new approach implies introduction of two different group transmission times -- the
exact group transmission time $\tau_{tr}^{ex}$ and the asymptotic group transmission
time $\tau_{tr}^{as}$. By \cite{Ch27}, the former is introduced as the difference
$t^{tr}_2-t^{tr}_1$, where $t^{tr}_1$ and $t^{tr}_2$ are such instants of time that
\begin{eqnarray*}
\fl
\frac{1}{\textbf{T}}\left(<\psi_{tr}(x,t^{tr}_1)|\hat{x}|\psi_{tr}(x,t^{tr}_1)>\right)=a;\ooo
\frac{1}{\textbf{T}}\left(<\psi_{tr}(x,t^{tr}_2)|\hat{x}|\psi_{tr}(x,t^{tr}_2)>\right)=b;
\end{eqnarray*}
here $<\psi_{tr}(x,t)|\hat{x}|\psi_{tr}(x,t)>=\int_{-\infty}^{x_c}
x\left|\Psi_{tr}(x,t)\right|^2dx+\int^{\infty}_{x_c}
x\left|\Psi_{full}(x,t)\right|^2dx$.

Analogously, for reflection the exact group time $\tau_{ref}^{ex}$ is defined as
$\tau_{ref}^{ex}=t^{ref}_2-t^{ref}_1$, where $t^{ref}_1$ and $t^{ref}_2$ are such
instants of time that
\begin{eqnarray} \label{219}
\fl\frac{1}{\textbf{R}}\left(<\psi_{ref}(x,t^{ref}_1)|\hat{x}|\psi_{ref}(x,t^{ref}_1)>\right)=
\frac{1}{\textbf{R}}\left(<\psi_{ref}(x,t^{ref}_2)|\hat{x}|\psi_{ref}(x,t^{ref}_2)>\right)=a.
\end{eqnarray}

As regards $\tau_{tr}^{as}$, it describes the influence of the potential barrier on
a particle within the wide spatial interval $[a-L_1,b+L_2]$ where $L_1,L_2\gg l_0\gg
d.$ In this case, instead of the exact wave functions for transmission, one can use
the corresponding incoming and outgoing waves,
\begin{eqnarray} \label{220}
\fl \psi_{tr}^{in,out}(x,t)=\frac{1}{\sqrt{2\pi}}\int_{-\infty}^{\infty}
\Api(k)f_{tr}^{in,out}(k)\exp[i(kx-E(k)t/\hbar)];
\end{eqnarray}
\begin{eqnarray*}
\fl f_{tr}^{in}(k,t)=\sqrt{T}\exp\left[i\left(\lambda
+sign(\lambda)\frac{\pi}{2}\right)\right],\ooa
f^{out}_{tr}(k)=\sqrt{T}\exp[i(J(k)-kd)];\ooa J=\arg(a_{out}).
\end{eqnarray*}

Long before and long after the scattering event the motion of the centroid of the
wave packet $\psi_{tr}(x,t)$ is described, respectively, by the expressions
\begin{eqnarray*}
\fl <\hat{x}>_{tr}^{in}=\frac{\hbar
t}{m}<k>_{tr}^{in}-<\lambda^\prime>_{tr}^{in};\ooa <\hat{x}>_{tr}^{out}=\frac{\hbar
t}{m}<k>_{tr}^{out}-<J^\prime>_{tr}^{out}+d;
\end{eqnarray*}
here $<k>_{tr}^{out}=<k>_{tr}^{in}=<k>_{tr}$ (see (\ref{444})); the brackets
$<\ldots>_{tr}^{in,out}$ denote averaging over the wave packets
$\psi_{tr}^{in,out}(x,t)$; the prime denotes the derivative with respect to $k$.

The time $\uta(L_1,L_2)$ spent by the centroid, located at the point
$<\hat{x}>_{tr}$, in the interval $[a-L_1,b+L_2]$ is
\begin{eqnarray*}
\fl \uta(L_1,L_2)\equiv t_{tr}^{(2)}-t_{tr}^{(1)}=\frac{m}{\hbar
<k>_{tr}}\left(<J^\prime>_{tr}^{out} -<\lambda^\prime>_{tr}^{in} +L_1+L_2 \right).
\end{eqnarray*}
The values of $t_{tr}^{(2)}$ and $t_{tr}^{(1)}$ obey the equations
\[\fl <\hat{x}>_{tr}^{in}(t_{tr}^{(1)})=a-L_1;\ooo <\hat{x}>_{tr}^{out}(t_{tr}^{(2)})=b+L_2.
\]

The term $\uta^{as}$ ($\uta^{as}=\uta(0,0)$) is just the asymptotic group
transmission time,
\begin{eqnarray} \label{230}
\fl \uta^{as}=\frac{m d^{eff}_{tr}}{\hbar <k>_{tr}},\ppp
d^{eff}_{tr}=<J^\prime>_{tr}^{out} -<\lambda^\prime>_{tr}^{in}.
\end{eqnarray}

Analogously, $\tau_{ref}(L_1)$ spent by the reflected centroid, located at the point
$<\hat{x}>_{ref}$, in the interval $[a-L_1,x_c]$ is
\begin{eqnarray*}
\fl\tau_{ref}(L_1)\equiv t^{ref}_2-t^{ref}_1=\frac{m}{\hbar
<k>^{ref}_{in}}\left(<J^\prime - F^\prime>^{ref}_{out}-<\lambda^\prime>^{ref}_{in}
+2L_1\right);
\end{eqnarray*}
here the instants of time $t^{ref}_1$ and $t^{ref}_2$ obey equations
\[\fl <\hat{x}>^{ref}_{in}(t^{ref}_1)=a-L_1,
\ppp<\hat{x}>^{ref}_{out}(t^{ref}_2)=a-L_1.
\]
Then the term $\utb^{as}$ ($\utb^{as}=\utb(0)$) is just the asymptotic group
reflection time,
\begin{eqnarray} \label{51250}
\fl\utb^{as}=\frac{m d^{eff}_{ref}}{\hbar <k>^{ref}_{in}},\ppp
d^{eff}_{ref}=<J^\prime-F^\prime>^{ref}_{out}-<\lambda^\prime>^{ref}_{in}
\end{eqnarray}
Parameters $d_{eff}^{tr}$ and $d_{eff}^{ref}$ can be interpreted as effective
barrier's widths associated with the transmission and reflection, respectively.
These quantities, together with asymptotic group scattering times $\uta^{as}$ and
$\utb^{as}$ can be negative by value, unlike exact group scattering times
$\uta^{ex}$ and $\utb^{ex}$.

Note that the average starting points $x^{start}_{tr}$ and $x^{start}_{ref}$ for
transmission and reflection, respectively, are determined by expressions
$x^{start}_{tr}=-<\lambda^\prime>^{tr}_{in}$ and
$x^{start}_{ref}=-<\lambda^\prime>^{ref}_{in}$. That is, they differ from
$x^{start}_{full}$ to characterize the whole ensemble of particles. This result
distinguishes our approach from the CMT based on the implicit assumption that
transmitted particles start, on the average, from the point $x^{start}_{full}$ (in
the setting considered, $x^{start}_{full}=0$ (see (\ref{444}))).

Let us consider in detail tunnelling a particle, with a given energy $E$, through
the rectangular potential barrier of height $V_0$ ($E\leq V_0$). Note, firstly, that
for symmetric potential barriers $F^\prime(k)\equiv 0$. Therefore
$d^{eff}_{tr}(k)=d^{eff}_{ref}(k)\equiv d_{eff}(k)$ and
$x^{start}_{tr}=x^{start}_{ref}\equiv x_{start}=-\lambda'(k)$. In this case
\[\fl\uta^{as}(k)=\utb^{as}(k)=\tau_{as}(k)=\frac{md_{eff}(k)}{\hbar k}.\]
\[\fl d_{eff}(k)=\frac{4}{\kappa}\cdot
\frac{\left[k^2+\kappa_0^2\sinh^2\left(\kappa d/2\right)\right]
\left[\kappa_0^2\sinh(\kappa d)-k^2 \kappa d\right]} {4k^2\kappa^2+
\kappa_0^4\sinh^2(\kappa d)};\]
\[\fl x_{start}(k)= -2\frac{\kappa_0^2}{\kappa}\cdot
\frac{(\kappa^2-k^2)\sinh(\kappa d)+k^2 \kappa d \cosh(\kappa d)} {4k^2\kappa^2+
\kappa_0^4\sinh^2(\kappa d)};\] here $\kappa_0=\sqrt{2mV_0}/\hbar$ (see also
(\ref{6})).

\begin{figure}[h]
\begin{center}
\includegraphics[width=8.0cm]{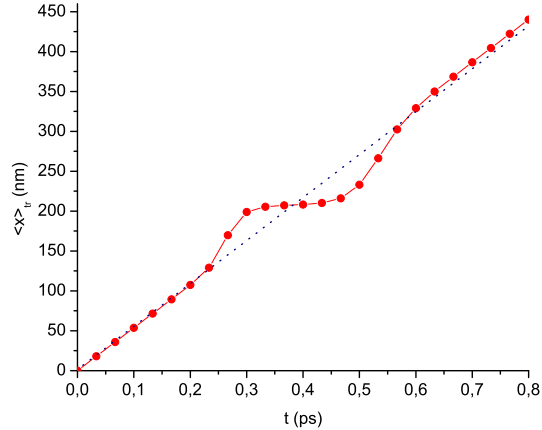}
\end{center}
\caption{The centroid's positions for $\psi_{tr}(x,t)$ (circles) and for the
corresponding freely moving wave packet (dashed line) as functions of time $t$.}
\label{fig:fig5a1}
\end{figure}

As is seen, like the phase time $\tau_{ph}$ defined in the STM for rectangular
barrier,
\begin{equation*}
\fl\tau_{ph}(k)= \frac{m}{\hbar k\kappa}\cdot \frac{2\kappa d
k^2(\kappa^2-k^2)+\kappa_0^4 \sinh(2\kappa d)}{4k^2\kappa^2+
\kappa_0^4\sinh^2(\kappa d)},
\end{equation*}
$\tau_{as}(k)$ saturates too with increasing the barrier's width $d$. However, this
fact does not at all mean that the effective velocity of a particle tunnelling
through a wide rectangular barrier becomes superluminal. It is demonstrated by Fig.
\ref{fig:fig5a1} which shows the function $<\hat{x}>_{tr}(t)$ to describe scattering
the Gaussian wave packet ($l_0=10nm$, $E_0=\hbar^2k_0^2/2m=0.05eV$) on the
rectangular barrier ($a=200nm$, $b=215nm$, $V_0=0.2eV$). (Note, in this case the
deviation of $\textbf{T}$ from $1-\textbf{R}$ does not exceed five percentages,
though the wave-packet's and barrier's widths are of the same order.)

This figure shows explicitly a qualitative difference between the exact and
asymptotic group times -- the latter is not an approximation of the former. While
the former gives the time spent by the centroid just in the barrier region (in the
CMT this role is played by $\tau_{ph}$), the latter describes the influence of the
barrier on the centroid in the course of the whole process. More precisely, the
quantity $\uta^{as}-\tau_{free}$, where $\tau_{free}=md/\hbar k_0$, is the time
delay to appear in the motion of the centroid of the transmitted wave packet, as
compared with that of the corresponding freely moving packet (it starts from the
point $x^{start}_{tr}$, rather than from $x^{start}_{full}$), in the course of a 1D
completed scattering. In the case considered, $\tau^{ex}_{tr}\approx 0,155ps$,
$\uta^{as}\approx 0,01ps$, $\tau_{free}\approx 0,025ps$.

As is seen, the influence of an opaque rectangular barrier on the transmitted wave
packet has a complicated character. The exact group time says that the centroid's
velocity inside the barrier region is much smaller than outside. While the
asymptotic group time tells us that the total influence of the barrier on the
transmitted wave packet has an accelerating character: the transmitted packet moves
ahead the corresponding packet moving freely. However, we have to take into account
that this effect is related to the asymptotically large spatial interval, as
$L_1,L_2\gg l_0\gg d.$ In this case the saturation of the asymptotic group
transmission time, with increasing the barrier's width, does not at all mean that
the centroid of the transmitted wave packet passes the barrier with a superluminal
velocity.

At the end of this section it is worthwhile to note that the effective width
$d_{eff}$ for the $\delta$-potential $V(x)= W \delta(x-a)$ is zero in our approach.
That is, the asymptotic group transmission time, like the exact one, is zero in this
case. This result is well expected for a point-like object to pass through the
barrier region of the $\delta$-potential, which is zero. Note that in this case
$x_{start}(k)=-2m\hbar^2 W/(\hbar^4 k^2+m^2W^2).$

Within the CMT we have an opposite situation. Now
$x_{start}(k)=x^{start}_{full}(k)=0$, but $d_{eff}=2m\hbar^2 W/(\hbar^4
k^2+m^2W^2)$. This result is usually explained by the nonlocality of tunnelling a
particle through $\delta$-potential. However, in our opinion, this explanation is
questionable, because it is based implicitly on the illegitimate substitution for a
particle by the wave packet to describe its state (more precisely, the state of the
corresponding ensemble of particles).

\subsection{The dwell and Larmor times for transmission and reflection}\label{a32}

Note, the concepts of the exact group transmission and reflection times are
considered, in our approach, as auxiliary ones. For example, it in fact unfit for
timing reflected particles in the limiting case $l_0\to \infty$. The point is that
the centroid of a too wide wave packet $\psi_{ref}(x,t)$ simply does not enter the
barrier region (Eqs. (\ref{219}) have no roots in this case). The main role in
timing a scattering particle in the barrier region is played here by the dwell and
Larmor times, closely connected with each other.

Remind that the dwell times are introduced in \cite{Ch27} for the stationary
scattering, i.e., for a particle with a given energy $E$. So, the transmission
($\tau^{tr}_{dwell}$) and reflection ($\tau^{ref}_{dwell}$) dwell times read as
\begin{eqnarray} \label{4005}
\fl \tau_{tr}^{dwell}(k)=\frac{1}{I_{tr}}\int_a^b|\psi_{tr}(x;k)|^2 dx \equiv
\frac{1}{I_{tr}}\int_a^{x_c}|\Psi_{tr}(x;k)|^2
dx+\frac{1}{I_{full}}\int_{x_c}^b|\Psi_{full}(x;k)|^2 dx
\end{eqnarray}
\begin{eqnarray} \label{5140014}
\fl\tau_{ref}^{dwell}(k)=\frac{1}{I_{ref}} \int_a^{x_c}|\psi_{ref}(x,k)|^2 dx\equiv
\frac{1}{I_{ref}} \int_a^{x_c}|\Psi_{ref}(x,k)|^2 dx;
\end{eqnarray}
where $I_{tr}=I_{full}=T(k)\hbar k/m$, $I_{ref}=R(k) \hbar k/m.$

The Larmor times introduced in \cite{Ch27} for the nonstationary case read as
\begin{eqnarray} \label{5190024}
\fl\tau^L_{tr}=\frac{1}{\textbf{T}}\int_{0}^{\infty}\varpi(k)T(k)\tau_{tr}^{dwell}(k)dk,\ppp
\tau^L_{ref}=\frac{1}{\textbf{R}}\int_{0}^{\infty}\varpi(k)R(k)\tau_{ref}^{dwell}(k)dk;
\end{eqnarray}
$\varpi(k)=|A^{in}(k)|^2-|A^{in}(-k)|^2$; for a {\it completed} scattering,
$|A^{in}(k_0)|\gg |A^{in}(-k_0)|.$

For the rectangular barrier, for $E<V_0$, we have
\begin{eqnarray} \label{4007}
\fl \tau_{tr}^{dwell}(k)=\frac{m}{2\hbar
k\kappa^3}\left[\left(\kappa^2-k^2\right)\kappa d +\kappa_0^2 \sinh(\kappa
d)\right];
\end{eqnarray}
\begin{eqnarray} \label{4008}
\fl\tau_{ref}^{dwell}(k)=\frac{m k}{\hbar \kappa}\cdot\frac{\sinh(\kappa d)-\kappa
d}{\kappa^2+\kappa^2_0 \sinh^2(\kappa d/2)}\ooo \mbox{äëÿ}.
\end{eqnarray}

In the CMT \cite{But}, the dwell time $\tau_{dwell}$ for rectangular barrier
($E<V_0$) reads as
\begin{equation} \label{4009}
\fl\tau_{dwell}(k)= \frac{m k}{\hbar \kappa}\cdot \frac{2\kappa d
(\kappa^2-k^2)+\kappa_0^2 \sinh(2\kappa d)}{4k^2\kappa^2+ \kappa_0^4\sinh^2(\kappa
d)}.
\end{equation}

As is seen from (\ref{4007}) and (\ref{4009}), unlike $\tau_{dwell}$ to appear in
the CMT, $\tau_{tr}^{dwell}$ increases exponentially rather than saturates in the
limit $d\to\infty.$ Thus, in our approach, both the exact group time and the dwell
time, contrary to the corresponding concepts introduced in the CMT, do not lead to
the Hartman effect. By our approach the opaque barrier strongly delays, on the
average, the motion of a particle when it enters the barrier region.

Moreover, as is shown in \cite{Ch29} for a particle tunnelling through the system of
two identical rectangular barriers of width $d$ and height $V_0$, with the distance
$l$ between them, this approach does not lead to the generalized Hartman effect
which was found within the CMT in \cite{Ol3}. In the opaque-barrier limit, when
$V_0\to\infty$ (or $k_0\to\infty$) and $d$ is fixed, for the the phase time
$\tau_{ph}$ and dwell time $\tau_{dwell}$ introduced in the CMT we have
\[\fl \tau_{ph}\approx \frac{2m}{\hbar kk_0},\ppp \tau_{dwell}\approx
\frac{2mk}{\hbar k_0^3}.\] As is seen, both these quantities do not depend on the
distance between the barriers and diminish when $k_0\to\infty$.

In order to show the behavior of $\tau_{tr}^{dwell}(k)$ and $\tau_{ref}^{dwell}(k)$
in this limit, let us firstly note that these quantities possess the property of
additivity. They can be written as follows, $\tau_{tr}^{dwell}=\tau^{(1)}_{tr}
+\tau^{gap}_{tr} +\tau^{(2)}_{tr}$ and $\tau_{ref}^{dwell}=\tau^{(1)}_{ref}
+\tau^{gap}_{ref}$; here $\tau^{(1)}_{tr}$ and $\tau^{(1)}_{ref}$ describe the first
barrier; $\tau^{gap}_{tr}$ and $\tau^{gap}_{ref}$ do the gap between the barriers,
and $\tau^{(2)}_{tr}$ describes the second barrier (remind that reflected particles
do not cross the midpoint of this symmetric structure). As is shown in \cite{Ch29},
in the above limit,
\[\fl\tau^{dwell}_{ref}\approx\tau^{(1)}_{ref}\approx \tau_{dwell},\ppp
\tau^{(1)}_{tr}=\tau^{(2)}_{tr}\approx\frac{m}{4\hbar kk_0}e^{2k_0d},\ppp
\tau^{gap}_{tr}\approx \frac{mk_0^2}{8\hbar k^4}\left(kl-\sin(kl)\right)e^{2k_0d}.\]
As is seen, $\tau_{tr}^{dwell}(k)$ increases exponentially in this limit, and what
is also important is that it depends on the distance $l$ between the barriers.
Moreover, in the course of passing a particle through the structure, it spends the
most part of time just in the space between the opaque barriers.

Analogous situation arises in another opaque-barrier limit, when $d\to\infty$ and
$V_0$ is fixed. Now the explicit expressions for the characteristic times becomes
somewhat complicated (see \cite{Ch29}), giving no qualitatively new information, and
we omit this case.

\section{On a strictly symmetrical setting of the double-slit experiment} \label{slit}

As is well known, the main puzzle of a double slit diffraction is that within its
conventional quantum model a particle possesses mutually exclusive properties - its
wave and corpuscular properties are incompatible with each other. But this result
cannot be considered as a finally established fact. Here we present a novel model of
a double slit diffraction where this process is treated, like a 1D completed
scattering, as a {\it compound} one-particle process. Within this model the wave and
corpuscular properties of a particle are in a peaceful coexistence with each other.

Of course, the way of decomposing this one-particle process into alternative
subprocesses is different. As is seen from Section \ref{a1}, the key role in
decomposing a 1D completed scattering into subprocesses is played by the fact that
at the final stage of the process the particle's state is a CSMDS in which substates
do not interfere with each other, being localized in the non-overlapped spatial
regions. Just the final substates to describe transmission and reflection were used
for reconstructing their whole evolution.

At first glance, in the case of a double-slit diffraction we meet a more comfortable
situation, because here the particle's state is a CSMDS at all stages of scattering.
However, the problem is that the norms of the one-slit substates to enter this
superposition do not give unit (the norm of the CSMDS) because of the interference
between them. This means that these (initially) distinct one-slit substates to
appear in the {\it one-slit} experiments associated with the first and second slits,
cannot be associated with the alternative subprocesses of the double-slit
experiment. Just this fact is usually interpreted as the incompatibility of the wave
and corpuscular properties of a particle in the double-slit experiment.

In this connection, our aim is to show, by the example of a strictly symmetrical
setting of the double-slit experiment, how to transform the above CSMDS, with the
interfering one-slit substates, into that with substates to describe alternative
subprocesses. It is assumed that the $YZ$-plane coincides with the plane of the
first screen to have two parallel identical slits centered on the planes $y=-a$
(first slit) and $y=a$ (second slit), and the wave function $\Psi_{two}(x,y,z,t)$ to
describe a particle when both the slits are opened has the form
\begin{equation} \label{71}
\fl\Psi_{two}(x,y,z,t)\equiv\Phi(x,y,z,t);
\end{equation}
\[\fl \Phi(x,y,z,t)=
\frac{1}{\sqrt{2}}\left[\Psi_{one}(x,y-a,z,t)+\Psi_{one}(x,y+a,z,t)\right]\] where
$\Psi_{one}(x,y,z,t)$ is the "one-slit" wave function; it is such that
$\Psi_{one}(x,-y,z,t)=\Psi_{one}(x,y,z,t)$, and $\Psi_{one}(0,y,z;k)=0$, if
$|y|>d/2$; $d$ is the slit's width; $a>d/2$. It is also assumed that a particle
impinges the first screen from the left, and the second screen -- the particle' s
detector -- coincides with the plane $x=L$ ($L>0$).

It is evident that in this case
\begin{equation} \label{81}
\fl\Psi_{two}(x,-y,z;k)=\Psi_{two}(x,y,z;k).
\end{equation}
Thus, the $y$-th components of the probability current densities associated with the
first and second slits balance each other on the plane $y=0$, and the $y$-th
component of the probability current density associated with their superposition is
zero on this plain (see also \cite{Hom} where this experiment is analyzed within the
Bohmian approach).

It is evident that, if to insert along the symmetry plane an infinitesimally thin
two-side "mirror"\ (we use here this word, bearing in mind the analogue between this
quantum experiment and its counterpart in classical electrodynamics) to elastically
scatter particles, the wave function (\ref{71}) and, hence, the interference pattern
to appear in this experiment on the second screen should remain the same; inserting
the mirror does not disturb the condition (\ref{81}).

The coincidence of the particle's states in the original (without the mirror) and
modified (with the mirror) settings of the double-slit experiment leads to the
following two conclusions: (\i) the ensemble of particles freely moving between the
first and second screens in the original setting is {\it equivalent} to the ensemble
of particles in the modified setting where they {\it a priori} cannot cross the
plane $y=0$ occupied by the mirror; (\i\i) in the original experiment a particle
always passes only through one of two open slits. Figuratively speaking, the above
procedure of inserting the mirror can be considered as a non-demolishing
"which-way"\ measurement.

From (\i) it follows that the wave function (\ref{71}) can be rewritten in the form
\begin{equation} \label{73}
\fl\Psi_{two}(x,y,z,t)=\psi_{two}^{(1)}(x,y,z,t)+\psi_{two}^{(2)}(x,y,z,t);
\end{equation}
where $\psi_{two}^{(1)}$ and $\psi_{two}^{(2)}$ are piecewise continuous wave
functions defined as follows,
\begin{eqnarray*} \fl\psi_{two}^{(1)}(x,y,z,t)=\left\{ \begin{array}{c}
0, \ppa y>0;\\
\Phi(x,y,z,t), \ooo y<0
\end{array}\right. \ppp
\psi_{two}^{(2)}(x,y,z,t)=\left\{ \begin{array}{c} \Phi(x,y,z,t),\ooo y>0;\\
0, \ppa y<0;
\end{array}\right.
\end{eqnarray*}
$\psi_{two}^{(1)}(x,0,z,t)=\psi_{two}^{(2)}(x,0,z,t)=\Phi(x,0,z,t)/2$. We have to
stress that these expressions are valid both for $x>0$ and for $x<0$. That is the
region of the particle's source is also divided by the symmetry plane into two
parts.

The substates $\psi_{two}^{(1)}$ and $\psi_{two}^{(2)}$ describe the alternative
subprocesses -- passing a particle through the first slit (provided that the second
one is opened) and passing a particle through the second slit (provided that the
first one is opened). These subprocesses are inseparable from each other -- either
subprocess creates an unremovable context for its counterpart.

Note, in this case, the second screen to play the role of the measurement device can
be considered as consisting of two (sub)devices -- the left half of the screen
($y<0$) to detect particles passing through the first slit, and the right half of
the screen ($y>0$) associated with the second slit. The same concerns the particle's
source located in the region $x<0$. So that particles to impinge upon the the left
(right) half of the second screen are emitted, in fact, from the left (right) half
of the particle's source.

So, though CSMDSs (\ref{71}) and (\ref{73}) give the same wave function, the
physical meaning of $\Psi_{one}(x,y-a,z,t)$ and $\Psi_{one}(x,y+a,z,t)$ to enter
(\ref{71}) differs cardinally from that of $\psi_{two}^{(1)}(x,y,z,t)$ and
$\psi_{two}^{(2)}(x,y,z,t)$ to enter (\ref{73}). The second pair describes the {\it
alternative} subprocesses of a double slit diffraction, because
$\|\Psi_{two}\|=\|\psi_{two}^{(1)}\|+ \|\psi_{two}^{(2)}\|$; here $\|\Psi\|$ stands
for $\langle\Psi|\Psi\rangle$. As regards the first pair, there are no alternative
subprocesses of a double slit diffraction which could be associated with this pair,
because $\|\Psi_{two}\|\neq\|\Psi_{one}(x,y-a,z,t)\|+ \|\Psi_{one}(x,y+a,z,t)\|$.

We have to stress that the following two one-particle processes -- passing a
particle through the first slit when the second one is {\it opened}, and passing a
particle through the first slit when the second one is {\it closed} -- have
different experimental contexts (see also Section \ref{context}). As a consequence,
they are described by different wave functions (see also \cite{Hr1}). By our
approach, $\psi_{two}^{(1)}(x,y,z,t)$ describes the first process, and
$\Psi_{one}(x,y+a,z,t)$ describes the second one.

Note, the "transformation" of the one-slit state $\Psi_{one}(x,y+a,z,t)$ into
$\psi_{two}^{(1)}(x,y,z,t)$, after opening the second slit (i.e., after transforming
the experimental context), results from the {\it interference} of the state
$\Psi_{one}(x,y+a,z,t)$ with $\Psi_{one}(x,y-a,z,t)$. Thus, the interference plays a
twofold role in this experiment. On the one hand, namely the interference makes it
impossible to associate the initial one-slit wave functions with alternative
subprocesses of a double slit diffraction. On the other hand, namely the
interference makes it possible to decompose a double slit diffraction into the
alternative subprocesses described by the inseparable one-slit substates
$\psi_{two}^{(1)}$ and $\psi_{two}^{(2)}$.

Note also that the CSMDS to describe a double slit diffraction, being written in the
form (\ref{73}), is similar to a mixed state, in the sense that it obeys the
"either-or"\ rule. Indeed, from (\ref{73}) it follows that a particle, despite (and,
simultaneously, due to) its wave properties, can pass {\it either} through the first
slit {\it or} through the second one -- the whole ensemble of particle consists of
two parts. However, unlike pure states to constitute a mixed state, the alternative
substates $\psi_{two}^{(1)}$ and $\psi_{two}^{(2)}$ to enter CSMDS (\ref{73}) are
inseparable from each other.

So, the main result of the model is that CSMDS (\ref{71}) with the initially
distinct one-particle substates, hidden due to the interference between them, has
been transformed into CSMDS (\ref{73}) with the inseparable one-slit substates,
distinct just due to the interference.

\section{The quantum-classical problem as that of reducing a non-Kolmogorovian
quantum probability space to underlie a CSMDS to the sum of classical ones}
\label{a4}

So, from the physical viewpoint the main innovation in a new wave-packet approach is
that it treats a 1D completed scattering and double slit diffraction as {\it
compound} processes, i.e., in the spirit of the classical physics. By this approach,
a micro-particle like a macro-particle can pass only through one of two open slits
in the screen, as well as it can {\it either} be transmitted {\it or} reflected by
the potential barrier, and so on. Thus, by this approach the abbreviation CSMDS is
in fact equally applicable both to {\it micro-} and {\it macro-}particles.

From the mathematical viewpoint its innovation is the representation of the wave
function to describe either of these two one-particle scattering processes as a
CSMDS whose unit norm equals to the sum of norms of its substates. That is in fact,
this approach represents the probability space associated with either process as the
sum of two probability subspaces associated with the subprocesses.

Our next step is to show that the presented here approach to the one-particle
scattering phenomena can be taken as the basis for the alternative programm of
resolving the long-standing quantum-classical problem, with keeping the linear
formalism of quantum mechanics and idealization of isolated systems.

As is known, by the quantum-classical problem is meant the conflict to arise, within
the contemporary quantum-mechanical description of CSMDSs, between the superposition
principle and the "either-or"\ rule to guide, in classical probability theory,
mutually exclusive random events (the conflict which is inadmissible for theory
pretending to the role of a universal one). The most important milestones in
developing the modern vision of this problem are the famous Schr\"odinger's cat
paradox and Bell's theory of the EPR-Bohm experiment. Both are aimed to demonstrate
in the most sharp form the existence of a deep contradiction between the quantum
laws of the micro-world and classical laws of the macro-world.

Figuratively speaking, on the road between the micro- and macro-scales,
Schr\"odinger and Bell go in the opposite directions. Schr\"odinger demonstrates the
appearance of the above conflict when one attempts to extend quantum laws onto the
macro-scales -- Schr\"odinger calls in question the validity of the superposition
principle at the macro-scales. While Bell, by developing the classical-like analysis
of the EPR-Bohm experiment, is aimed to show the appearance of this conflict when
one attempts to apply classical laws onto the micro-scales -- Bell calls in question
the universal validity of such fundamental notion of classical physics as "causal
external world", i.e., the world whose existence is independent of an observer, and
which is guided by the principles of special relativity.

Note that, since quantum mechanics has been developed by its founders as a {\it
universal} theory, the quantum-classical problem can be also interpreted as the
problem of the (in)completeness of quantum mechanics. From this viewpoint,
Schr\"odinger calls in question the {\it completeness} of quantum mechanics at the
macro-scales, and Bell, on the contrary, calls in question its {\it incompleteness}
at the micro-scales.

At present the most of scientists to deal with the quantum-classical problem treats
it as the measurement (or macro-objectification) problem whose resolution is
impossible within the idealization of isolated systems. This viewpoint is based on
the widespread interpretation of the experimentally observed violation of Bell's
inequalities. It is considered that their experimental violation falsifies the
assumption on the existence of local hidden variables (or other assumptions to
concern the (non)locality and/or (non)reality of the micro-world) to underlie Bell's
inequalities.

Within this picture, solving the quantum-classical problem to arise for CSMDSs is
impossible without suggesting that there is "everything else" (e.g., the mechanisms
of decoherence or localization) to influence (together with the considered
potential) the dynamics of the system under study, reducing its original state - a
CSMDS - to one of its microscopically distinct substates (the review of the
approaches based on this idea, which constitute the decoherence and
spontaneous-localization programs of solving the measurement (or
macro-objectification) problem, is done in \cite{Sch}; see also \cite{Joo,Ghi}).

Undoubtedly, extending quantum theory onto open micro-systems is of importance,
because there are many interesting physical problems when the influence of external
factors on a micro-system is essential. It should be stressed also that the deep
analysis of the foundations of quantum mechanics, which was carried out within these
programs, is important for searching for the possible ways of resolving the
quantum-classical problem. However the ways of resolving this problem, presented in
these programs themselves, are unacceptable, as they create a clearly deadlock
situation in physics. In fact they condemn quantum mechanics, with its inherent
superposition principle, as a weak theory being unable to describe the macro-world
and leaving unspeakable the micro-world.

In this connection, of importance are approaches to have tested the validity of
Bell's proof of the nonexistence of local hidden variables, from the mathematical
viewpoint. For example an important aspect of this question is pointed out in
\cite{Vo1,Vo2} to argue that in order to judge on (non)locality in the thought EPR
and EPR-Bohm experiments, one has at least to introduce a correct space-time
structure into their mathematical models. As is shown, this step is sufficient for
explaining the (original) EPR experiment.

Other arguments against the nonexistence of local hidden variables have been
developed within the approaches to reveal, from the viewpoint of quantum mechanics
and classical probability theory, all assumptions to underlie the classical-like
derivation of Bell's inequalities (see the pioneer works by Fine \cite{Fin},
Pitowsky \cite{Pit} and Accardi \cite{Acc}, as well as the recent review \cite{Hre}
and papers \cite{Hes,Nie,An1,Joy}).

As was shown by Fine \cite{Fin}, apart from the explicit assumption on the existence
of local hidden variables, the derivation of Bell's inequality is based also on the
implicit assumption that there is a compatible joint distribution to describe
experimental data obtained in the EPR-Bohm experiment for different orientations of
particle's detectors. From the viewpoint of quantum mechanics this assumption is
improper {\it a priori}, because such data are obtained in fact for noncommuting
observables and, thus, there is no compatible joint quantum distribution to describe
them. That is, it is not surprising that quantum probabilities violate this
inequality.

However, this assumption is also at variance with classical probability theory (see
\cite{Pit,Acc}) where Bell's type inequalities have been known yet before Bell's
theorem, and their violation means simply that the probabilities to enter these
inequalities are incompatible -- they describe statistical (experimental) data which
cannot be associated with a common Kolmogorovian probability space. A detailed
analysis of Bell's theorem, from the viewpoint of probability theory, is done in the
studies reviewed in \cite{Hre}.

As regards the peculiarities of its experimental testing, as was stated in
\cite{An1}, "Strictly speaking, there does not exist [Bell's] inequality such that
all the three means involved in it would be spin correlations. It is therefore
meaningless to speak of verification of [this] inequality\ldots". Local hidden
variables and probabilities, consistently introduced for the experiments to test
Bell's inequality, obey the inequality to differ from Bell's one [ibid].

Thus, the experimental violation of Bell's inequality does not falsify the existence
of local hidden variables. Rather it falsifies the existence of "Bell's local hidden
variables"\ introduced inconsistently. The main lesson to follow from the critique
of Bell's theorem is that quantum mechanics and classical probability theory respect
each other in describing the EPR-Bohm experiment -- introducing a {\it common} local
hidden variable for different experimental contexts contradicts both these theories.

The above studies to revise the role of Bell's inequalities are of great importance
for solving the quantum-classical problem, because they, {\it in part}, rid of
obstacles the road from the macro- to the micro-world, for local hidden variables.
(We have to stress that it is also the great service of Bell who introduced these
inequalities for the analysis of this problem.) However, they leave untouched the
contradiction between the superposition principle and "either-or" rule to guide
local hidden variables. As before, the conventional quantum-mechanical description
of CSMDSs prevents extending local hidden variables onto the micro-level and, vice
versa, extending the superposition principle onto the macro-level -- the
Schr\"odinger cat paradox has remained unresolved.

In fact we return in the epoch preceding Bell's one. Again, due to the conventional
description of CSMDSs, the standard quantum mechanics based on the idealization of
isolated systems looks as theory complete at the micro-level but incomplete at the
macro-level. Thus, the presented in \cite{Sch} programs of solving the
quantum-classical problem, which aimed to make this theory complete at the
macro-level, remain relevant as before.

At the same time, the above models show that the conventional description of CSMDSs
is not a finally established fact. In studying such states, we have to take into
account that a CSMDS, as the EPR-Bohm experiment, deals with incompatible
statistical data. Indeed, this pure state implies that the system under study
evolves under some {\it complex} experimental context consisting of several {\it
elementary} contexts to imply alternative variants of evolution. For example, in the
case of a 1D completed scattering, two elementary contexts are associated with two
detectors, for transmitted and reflected particles; in the case of a double slit
diffraction, either of two slits creates its own (elementary) experimental context.

Of importance is to stress that the elementary contexts are integral parts of the
whole (complex) experimental context to be the "calling card"\ of the phenomenon
under study. This means that the elementary contexts are inseparable from each
other, what, in its turn, results in the inseparability of the corresponding
subprocesses.

From the viewpoint of classical probability theory, a CSMDS describes statistical
data to belong to a {\it non-Kolmogorovian} probability space. This was shown
explicitly in \cite{Ac1} by the example of a double slit diffraction. According to
classical probability theory, each quantum phenomenon described by a CSMDS should be
treated as a compound process to consist of several alternative subprocesses. This
implies that the non-Kolmogorovian probability space to underlie a CSMDS should be
reduced to the sum of Kolmogorovian ones associated with the subprocesses.

What is important is that quantum mechanics needs the same! Indeed, the above
studies of the EPR-Bohm experiment teach us that the non-Kolmogorovness of some
model of a quantum phenomenon means that it deals in fact with noncommuting
observables. That is, there is no observable which could be associated with
incompatible statistical data described by a CSMDS. In particular, introducing any
chracteristic time as well as calculating the expectation value of any one-particle
observable over the whole ensemble of particles described by a CSMDS is meaningless.

So, we arrive at the following two conclusions: (\i) the contemporary description of
CSMDSs is contradictory; (\i\i) the quantum-classical problem should be considered
as a purely quantum-mechanical problem, namely as that of modelling the quantum
phenomenon described by a CSMDS as a compound one consisting of alternative
subprocesses, or as that of reducing a non-Kolmogorovian quantum probability space
to underlie a CSMDS to the sum of classical ones.

The presented here approach to a 1D completed scattering and double slit diffraction
just demonstrates how to resolve this problem on the basis of the standard
Schr\"odinger equation. As was shown, this cannot be done only for wide (in the
momentum space) wave packets scattering on a 1D potential barrier, at the very stage
of scattering. However, this fact does not prevent the quantum-to-classical
transition, because for the initial and final stages of a 1D completed scattering,
when the distances between the wave packets and potential barrier are asymptotically
large, the subprocesses can be distinguished in the general case too.

So, by our approach the "either-or"\ rule holds for all coherent superpositions of
distinct substates, both for micro- and macro-particles. That is, in the
Schr\"odinger's cat paradox, the long-suffering cat is {\it either} died {\it or}
alive, independently of an observer, because at any instant of time the radioactive
nucleus {\it either} has already decayed {\it or} has yet non-decayed.

\section{Some remarks on the concepts of reality and contextuality in a new
wave-packet approach} \label{context}

There is a widespread viewpoint that, within the statistical interpretation of
quantum mechanics, the one-particle wave function (to describe a particle in the
infinite number of identical experiments) corresponds to nothing in the physical
world. A new wave-packet approach justifies and simultaneously falsifies such a
viewpoint.

Indeed, on the one hand, any averaging over the CSMDSs to describe a 1D completed
scattering and double slit diffraction is indeed meaningless -- for either of these
phenomena, there is no one-particle observable which could be introduced for the
whole ensemble of particles described by the corresponding CSMDS, because it deals
with incompatible statistical data. On the other hand, such observables can be
introduced for their substates associated with Kolmogorovian probability subspaces.

For these phenomena, every one-particle observable splits into two obserables for
the inseparable subprocesses. This means that in the case of a CSMDS every
observable must be endowed with the additional index to specify the corresponding
subprocess. So, in the case of a 1D completed scattering the particle's position $x$
and momentum $p$ split, respectively, into the pairs ($x_{tr}$, $x_{ref}$) and
($p_{tr}$, $p_{ref}$). Analogously, in the case of a double slit diffraction we have
the pairs ($x_{two}^{(1)}$, $x_{two}^{(2)}$) and ($p_{two}^{(1)}$, $p_{two}^{(2)}$).
Quantum mechanics implies the existence of local hidden variables, and the wave
functions of subprocesses give the one-particle distributions
($x_{tr}$-distribution, $\dots$ $p_{two}^{(2)}-$distribution) to reflect
statistically the inherent properties of a (micro)particle, as a local entity.

We have to stress once more that both the models are non-contextual in the sense
that they imply the dependence of all one-particle variables only on the potential
under which a particle moves. They do not imply any dependence of the particle's
dynamics on some external factor, environment or an observer. At the same time the
subprocesses are contextual. As was said above, each subprocess creates an
unremovable context for its counterpart.

Note, in both models the contextuality of subprocesses is associated with symmetry.
So, the context to arise for each subprocess of a 1D completed scattering is
different in the spatial regions to lie on the different sides of the point $x_c$,
because, from the classical viewpoint, reflected particles cannot {\it a priori}
cross the midpoint $x_c$ of a symmetric potential barrier. It is the main reason why
each subprocess is described in these regions by different (properly matched at the
point $x_c$) solutions of the Schr\"odinger equations.

Similarly, the symmetry of the considered setting of the double slit experiment
leads to the piecewise continuous wave functions $\psi_{two}^{(1)}(x,y,z,t)$ and
$\psi_{two}^{(2)}(x,y,z,t)$ to describe the subprocesses evolving on the different
sides of the symmetry plane $y=0$. Inserting the two-side mirror along this plane
changes the experimental context. However it keeps the symmetry of the original
double slit experiment. As a consequence, the original experiment, where particles
move freely between the screens, and the modified ("which-way") experiment, with the
mirror inserted between the screens, are described by the same wave function.

\section{Conclusion}

So, we have developed a new approach to a 1D completed scattering and double slit
diffraction, by which these two one-particle processes are modelled as compound ones
consisting of two alternative inseparable subprocesses (each subprocess creates an
unremovable context for its counterpart). As was shown by Accardi, by the example of
a double slit diffraction, the probability spaces to underlie CSMDSs are
non-Kolmogorovian. Thus, from the viewpoint of probability theory, our approach
reduces a non-Kolmogorovian quantum probability space to the sum of two
Kolmogorovian (classical) subspaces.

Since a non-Kolmogorovian probability space describes incompatible statistical data,
there are no one-particle observable which could be defined on this space. In the
case of a CSMDS consisting of two alternative substates, any observable splits into
two noncommuting observables. Calculating the expectation value of any physical
observable as well as introducing characteristic times for a particle described by a
CSMDS are meaningful only for its substates. Just ignoring this feature, in the
current description of such pure states, leads to paradoxical effects (e.g., Hartman
effect and simultaneous passage of a particle through two slits in the screen).

Our models give the basis for resolving the quantum-classical problem. By them, a
micro-object described by a CSMDS is guided by the "either-or"\ rule. In this case,
the abbreviation CSMDS is equally applicable to micro- and macro-particles. Of
course, a complete resolution of the quantum-classical problem implies revising all
current models of quantum phenomena where CSMDSs appear. All they should be treated
as compound phenomena consisting of alternative inseparable subprocesses.

\section{Acknowledgments}

The author thanks V.G. Bagrov and Erasmo Recami for useful discussions of some
aspects of the paper. This work was supported (in part) by Russian Science and
Innovations Federal Agency under contract No 02.740.11.0238 as well as by the
Programm of supporting the leading scientific schools of RF (grant No 3558.2010.2).


\section*{References}
\begin{thebibliography}{861}

\bibitem{Leg}
Leggett, A. J.: Testing the limits of quantum mechanics: motivation, state of play,
prospects. J. Phys.: Condens. Matter. 14, R415-R451 (2002)
\bibitem{Bell}
Bell, J. S.: Speakable and unspeakable in quantum mechanics. Cambridge (2004)
\bibitem{Hre}
Khrennikov, A. Yu.: EPR-Bohm experiment and Bell's inequality: Quantum physics meets
probability theory. Theor. and Math. Phys. 157, 99-115 (2008); arXiv:0709.3909v2
\bibitem{Ac1}
Accardi, L.: Snapshots on quantum probability. Vestnik of Samara State University.
Natural Science Series 67, No 8/1, 277-294 (2008)
\bibitem{Ch26}
Chuprikov, N. L.: New approach to the quantum tunnelling process: Wave functions for
transmission and reflection. Russian Physics Journal. 49, 119-126 (2006)
\bibitem{Ch27}
Chuprikov, N. L.: New approach to the quantum tunnelling process: Characteristic
times for transmission and reflection. Russian Physics Journal. 49, 314-325 (2006)
\bibitem{Ch3}
Chuprikov, N. L.: On a new mathematical model of tunnelling. Vestnik of Samara State
University. Natural Science Series. 67, No 8/1, 625-633 (2008)
\bibitem{Ch29}
Chuprikov, N. L.: On the generalized Hartman effect: Scattering a particle on two
identical rectangular potential barriers.  arXiv:1005.1323v2
\bibitem{Bohr}
Bohr, B.: Maxwell and modern theoretical physics. Nature 128, 691-692 (1931)
\bibitem{Ha2}
Hauge, E. H. and St\o vneng, J. A.: Tunneling times: a critical review. Rev. of Mod.
Phys. 61, 917-936 (1989)
\bibitem{La1}
Landauer, R. and Martin, Th.: Barrier interaction time in tunnelling. Rev. Mod.
Phys. 66, 217-228 (1994)
\bibitem{Olk}
Olkhovsky, V. S. and Recami, E.: Recent developments in the time analysis of
tunnelling processes. Physics Reports. 214, 339-356 (1992)
\bibitem{Ste}
Steinberg, A. M. and Chiao, R. Y.: Tunneling delay times in one and two dimensions.
Phys. Rev. A49, 3283-3295 (1994)
\bibitem{Mu0}
Muga, J. G., Leavens, C. R.: Arrival time in quantum mechanics. Physics Reports.
338, 353-438 (2000)
\bibitem{Nu0}
Carvalho, C. A. A., Nussenzveig, H. M.: Time delay. Physics Reports. 364, 83-174
(2002)
\bibitem{Ol2}
Olkhovsky, V. S., Recami, E., Jakiel, J.: Unified time analysis of photon and
particle tunnelling. Physics Reports. 398, 133–178 (2004)
\bibitem{Win}
Winful, H. G.: Tunneling time, the Hartman effect, and superluminality: A proposed
resolution of an old paradox. Physics Reports. 436, 1-69 (2006)
\bibitem{Har}
Hartman, T. E.: Tunneling of a Wave Packet. J. Appl. Phys. 33, 3427-3433 (1962)
\bibitem{But}
Buttiker, M.: Larmor precession and the traversal time for tunnelling. Phys. Rev. B.
27, 6178-6188 (1983)
\bibitem{Mu6}
Muga, J. G., Egusquiza, I. L., Damborenea, J. A., Delgado, F.: Bounds and
enhancements for negative scattering time delays. Phys. Rev. A. 66, 042115(1-8)
(2002)
\bibitem{Wi1}
Winful, H. G.: Delay Time and the Hartman Effect in Quantum Tunneling. Phys. Rev.
Lett. 91, P.260401(1-4) (2003)
\bibitem{Ol1}
Olkhovsky, V. S., Petrillo, V. and Zaichenko, A. K.: Decrease of the tunnelling time
and violation of the Hartman effect for large barriers. Phys. Rev. A. 70,
034103(1-4) (2004)
\bibitem{So1}
Sokolovski, D., Msezane, A. Z., Shaginyan, V. R.: "Superluminal"\ tunnelling as a
weak measurement effect. Phys. Rev. A. 71, 064103(1-4) (2005)
\bibitem{Rec}
Recami, E. and Zamboni-Rached, M.: Localized Waves: A Review. In: Hawkes, P. W.
(ed.) Advances in Imaging and Electron Physics, vol. 156, pp. 235-353. San Diego,
Academic Press (2009)

\bibitem{Krek}
Krekora, P., Su, Q. and Grobe, R.: Effects of relativity on the time-resolved
tunnelling of electron wave packets. Phys. Rev. A. 63, 032107(1-8) (2001)
\bibitem{Zhi}
Li, Z. J. Q., Nie, Y. H., Liang, J. J. and Liang, J. Q.: Larmor precession and dwell
time of a relativistic particle scattered by a rectangular quantum well. J. Phys. A:
Math. Gen. 36, 6563-6570 (2003)
\bibitem{Lun}
Lunardi, J. T. and Manzoni, L. A.: Relativistic tunnelling through two successive
barriers. Phys. Rev. A. 76, 042111(1-6) (2007)
\bibitem{Nim}
Nimtz, G.: On Virtual Phonons, Photons, and Electrons. Found. Phys. DOI
10.1007/s10701-009-9356-z (2009)
\bibitem{La2}
Buttiker, M. and Landauer, R.: Traversal Time for Tunneling. Phys. Rev. Lett. 49,
1739-1742 (1982)

\bibitem{Hr1}
Khrennikov, A. Yu.: Interpretations of probability. VSP, Utrecht (1999)

\bibitem{Ol3}
Olkhovsky, V. S., Recami E. and Salesi G.: Superluminal tunnelling through two
successive barriers. Europhys. Lett. 57, 879-884 {2002}.
\bibitem{Hom}
Home, D. and Kaloyerou, P. N.: New twists to Einstein's two-slit experiment:
complementarity vis-$\grave{a}$-vis the causal interpretation. J. Phys. A: Math.
Gen. 22, 3253-3266 (1989)
\bibitem{Sch}
Schlosshauer, M.: Decoherence, the measurement problem, and interpretations of
quantum mechanics. Rev. of Mod. Phys. 76, 1268-1305 (2004)
\bibitem{Joo}
Joos E. and Zeh H. D.: The emergence of classical properties through interaction
with the environment. Z. Phys. B 59, 223-243 (1985)
\bibitem{Ghi}
Ghirardi, G. C., Grassi R, and Rimini A.: Continuous-spontaneous-reduction model
involving gravity. Phys. Rev. A 42, 1057–1064 (1990)

\bibitem{Vo1}
Volovich, I. V.: Quantum cryptography in spase and Bell's theorem. In: Khrennikov A.
Yu., (ed.) Proc. Conf. Foundations of Probability and Physics. Ser. Quantum
Probability and While Noise Analysis, vol. 13, pp. 364-372. World Sci., River Edge,
NJ (2001)
\bibitem{Vo2}
Volovich, I. V.: Towards Quantum Information Theory in Space and Time. In:
Khrennikov A. Yu. (ed.) Proc. Conf. Quantum Theory: Reconsideration of Foundations,
Ser. Math. Modelling. Vaxjo Univ. Press. vol. 2, pp. 423-440 (2002)

\bibitem{Fin}
Fine, A.: Hidden Variables, Joint Probability, and the Bell Inequalities. Phys. Rev.
Lett. 48, 291-295 (1982)
\bibitem{Pit}
Pitowsky, I.: Resolution of the Einstein-Podolsky-Rosen and Bell paradoxes. Phys.
Rev. Lett. 48, 1299-1302 (1982)
\bibitem{Acc}
Accardi, L.: The probabilistic roots of the quantummechanical paradoxes, in: The
Wave-Particle Dualism, Diner, S. et al. (eds.), D. Reidel, Dordrecht (1984)

\bibitem{Hes}
Hess, K., Michielsen, K. and De Raedt, H.: Possible experience: From Boole to Bell.
EPL. 87, No 6, 60007(1-6) (2009)

\bibitem{Nie}
Nieuwenhuizen, T.M.: Is the Contextuality Loophole Fatal for the Derivation of Bell
Inequalities? Found Phys DOI 10.1007/s10701-010-9461-z

\bibitem{An1}
Andreev, V. A.: The correlation Bell inequalities. TMF. 158:2, 234–249 (2009)
\bibitem{Joy}
Christian, J.: Disproofs of Bell, GHZ, and Hardy Type Theorems and the Illusion of
Entanglement. arXiv:0904.4259 (2009)



\end{thebibliography}
\end{document}